\documentclass[%
 amsmath,amssymb,
 aps,
prl, reprint,
]{revtex4-1}

\usepackage{placeins}
\usepackage{graphicx}
\usepackage{bm}

\begin{document}

\preprint{APS/123-QED}

\title{Probing quantum capacitance in a 3D topological insulator}

\author{D. A. Kozlov$^{1,2,3}$, D. Bauer$^{3}$, J. Ziegler$^{3}$, R. Fischer$^{3}$, M. L. Savchenko$^{1,2}$, Z. D. Kvon$^{1,2}$, N. N. Mikhailov$^{1}$, S. A. Dvoretsky$^{1}$, D. Weiss$^{3}$}.
\affiliation{$^1$A. V. Rzhanov Institute of Semiconductor Physics,
Novosibirsk 630090, Russia}
\affiliation{$^2$Novosibirsk State University, Novosibirsk,
630090, Russia}
\affiliation{$^3$Experimental and Applied Physics, University of
Regensburg, D-93040 Regensburg, Germany}

\date{\today}

\begin{abstract}
We measure the quantum capacitance and probe thus directly the electronic density of states of the high mobility, Dirac type of two-dimensional electron system, which forms on the surface of strained HgTe. Here we show that observed magneto-capacitance oscillations probe – in contrast to magnetotransport - primarily the top surface. Capacitance measurements constitute thus a powerful tool to probe only one topological surface and to reconstruct its Landau level spectrum for different positions of the Fermi energy.

\begin{description}
\item[PACS numbers]
\pacs{1} 73.25.+i, \pacs{2} 73.20.At, \pacs{3} 73.43.-f
\end{description}
\end{abstract}

\pacs{1}
\maketitle

Three dimensional topological insulators (3D TI) represent a new class of materials with insulating bulk and conducting two-dimensional surface states \cite{Hasan10, Moore10, Xiao11, Ando13}. The properties of these surface states are of particular interest as they have a spin degenerate, linear Dirac like dispersion with spins locked to their electron's $k$-vector \cite{Ando13,Kane05-2}. Strained epilayers of HgTe, examined here, constitute a 3D TI with high electron mobilities allowing the observation of Landau quantization and quantum Hall steps down to low magnetic fields \cite{Brune11, Kozlov14}. While unstrained HgTe is a zero gap semiconductor with inverted band structure \cite{Groves67, Seeger85}, the degenerate $\Gamma 8$ states split and a gap opens at the Fermi energy $E_F$ if strained. This system is a strong topological insulator \cite{Kane07}, explored so far by transport \cite{Brune11, Kozlov14, Brune14},  ARPES \cite{Crauste13}, photoconductivity and magneto-optical experiments \cite{Shuvaev12, Shuvaev13, ShuvaevPRB13, DantscherPRB14}; also the proximity effect at superconductor/HgTe has been investigated \cite{Sochnikov15}. Since these two-dimensional electron states (2DES) have high electron mobilities of several $10^5$\,cm$^2$/V$\cdot$s, pronounced Shubnikov- de Haas (SdH) oscillations of the resistivity and quantized Hall plateaus commence in quantizing perpendicular magnetic fields \cite{Brune11,Kozlov14,Brune14}, stemming from both, top and bottom 2DES. The origin of the oscillatory resistivity is Landau quantization which strongly modifies the density of states (DoS). Capacitance spectroscopy allows to directly probe the thermodynamic DoS $dn/d\mu$ ($n$ = carrier density, $\mu$ = electrochemical potential) denoted as $D$, of a 3D TI. The total capacitance measured between a metallic top gate and a two-dimensional electron system (2DES) depends, besides the geometric capacitance, also on the quantum capacitance $e^2 D$ , connected in series and reflecting the finite density of states $D$ of the surface electrons \cite{Luryi88, Smith86, Mosser86}; $e$ is the elementary charge. Below, the quantum capacitance of the top surface layer is denoted as $e^2 D_{\rm t}$, the one of the bottom layer by $e^2 D_{\rm b}$.
\begin{figure}[t]
\includegraphics[width=0.745\columnwidth,keepaspectratio]{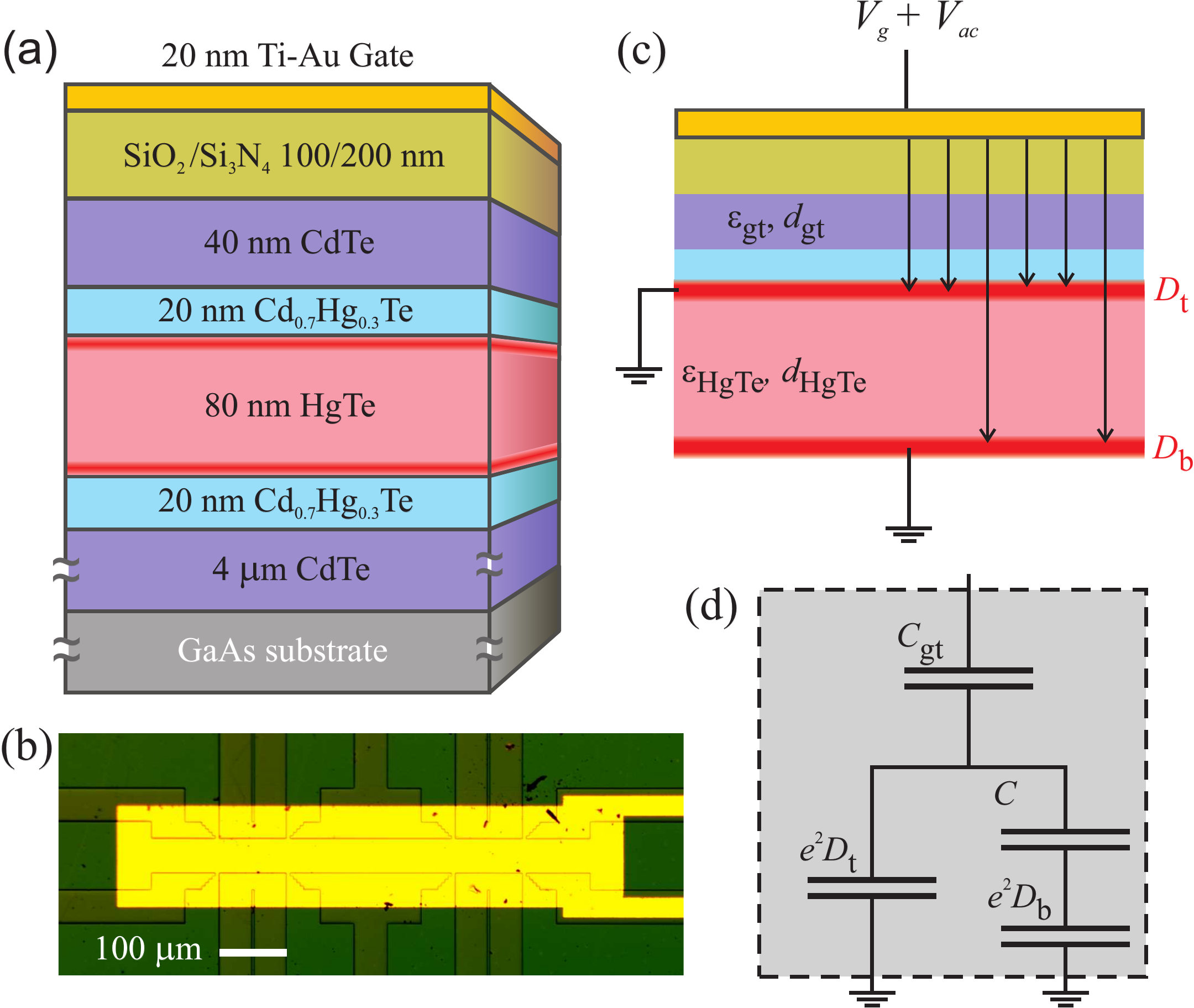}
\caption{\label{fig1} (a) Schematic cross section of the heterostructures studied. The Dirac surface states, shown in red, enclose the strained HgTe layer. (b) Top view of the device showing the Hall geometry with 8 potential probes at the side, covered by the top gate (yellow). (c) Schematic illustration of the ``three plate capacitor'' formed by the metallic top gate electrode and top and bottom layer (red) of the two-dimensional Dirac surface electrons with density of states $D_{\rm t}$ and $D_{\rm b}$ of top and bottom layer, respectively. The geometric capacitance $C_{\rm tb} = \varepsilon_{\rm HgTe} \varepsilon_0 A / d_{\rm HgTe}$ is determined by the gate area $A$, dielectric constant and thickness of CdTe, SiO$_2$ and Si$_3$N$_4$ layers. The electric field (black arrows) is partially screened by the top surface layer. (d) Equivalent circuit of the three plate capacitor with the quantum capacitances $A e^2 D_{\rm t}$ and $A e^2 D_{\rm b}$  in series with the respective geometrical capacitances. The equivalent circuit is the one introduced in \cite{Luryi88}, but extended by the quantum capacitance of the bottom surface, $A e^2 D_{\rm b}$. A similar equivalent circuit was recently introduced in \cite{Baum14}.}
\end{figure}

The experiments are carried out on strained 80\,nm thick HgTe films, grown by molecular beam epitaxy on CdTe (013) \cite{DantscherPRB14}. The Dirac surface electrons have high electron mobilities of order $4\times10^5$\,cm$^2$/V$\cdot$s. The cross-section of the structure is sketched in Fig.~1a. For transport and capacitance measurements, carried out on one and the same device, the films were patterned into Hall bars with metallic top gates (Fig.~1b).  For gating, two types of dielectric layers were used, giving similar results: 100\,nm SiO$_2$ and 200\,nm of Si$_3$N$_4$ grown by chemical vapor deposition or 80\,nm Al$_2$O$_3$ grown by atomic layer deposition. In both cases, TiAu was deposited as a metallic gate. The measurements were performed at temperature  $T=1.5$\,K and in magnetic fields $B$ up to 13\,T. Several devices from the same wafer have been studied. For magnetotransport measurements standard lock-in technique has been applied with the excitation AC current of 10-100\,nA and frequencies from 0.5 to 12\,Hz. For the capacitance measurements we superimpose the dc bias voltage $V_g$ and a small ac bias voltage (see Fig.~1c) and measure the ac current flowing across our device phase sensitive using lock-in technique. The typical ac voltage was 50\,mV at a frequency of 213\,Hz. The absence of both leakage currents and resistive effects in the capacitance were controlled by the real part of the measured ac current. In order to avoid resistive effects in high magnetic fields additional measurements at lower frequencies (up to several Hz) were performed.

When the Fermi level (electrochemical potential) is located in the bulk gap the system can be viewed as a "three-plate" capacitor where the top and bottom surface electrons form the two lower plates (see Fig.~1c and the corresponding equivalent circuit in Fig.~1d). From this equivalent circuit follows that, as long as $D_b$ does not vanish, the measured total capacitance is more sensitive to changes of $D_{\rm t}$ than of $D_{\rm b}$; the explicit connection between $D_t$, $D_b$ and the total capacitance $C$ is given in the Supplemental Material. The ratio of $(dC/dD_{\rm t}) / (dC/dD_{\rm b})=\Bigl( A D_{\rm b} e^2 + C_{\rm tb} \Bigr)^2 / C_{\rm tb}^2$ is significantly larger than unity since $A D_{\rm b} e^2 + C_{\rm tb}$ is (at $B = 0$) at least a factor of two larger than $C_{\rm tb}$ . Here, $C_{\rm tb}$ is the geometric capacitance $C_{\rm tb} = \varepsilon_{\rm HgTe} \varepsilon_0 A / d_{\rm HgTe}$ between top and bottom layer, with  $\varepsilon_{\rm HgTe} \approx 21$ \cite{Baars72}, the dielectric constant of HgTe, $d_{\rm HgTe}$ is the thickness of the HgTe film, $A$ is the gated TI area and $\varepsilon_0$ the dielectric constant of vacuum. Therefore, the measured capacitance reflects primarily the top surface's DoS, $D_{\rm t}$. In the limit $C_{\rm tb} \rightarrow 0$ (infinite distance to bottom surface) or  ($e^2 D_{\rm b}\rightarrow 0$) (no charge on bottom surface) the total capacitance $C$ is given by the expression usually used to extract the DoS $D$ of a two-dimensional electron system: $1/C = 1/C_{\rm gt} + 1/A e^2 D$  with the geometric capacitance $C_{\rm gt}=\varepsilon_{\rm gt} \varepsilon_0 A / d_{\rm gt}$, where $\varepsilon_{\rm gt}$ is the dielectric constant of the layers between gate and top 2DES, and $d_{\rm gt}$, is the corresponding thickness \cite{Smith86,Mosser86}. Note that $C_{\rm gt} \ll e^2 D_{\rm t}$ therefore the variations of the DoS cause only small changes of the measured value of $C$.

\onecolumngrid

\begin{figure}[t]
\includegraphics[width=1\columnwidth,keepaspectratio]{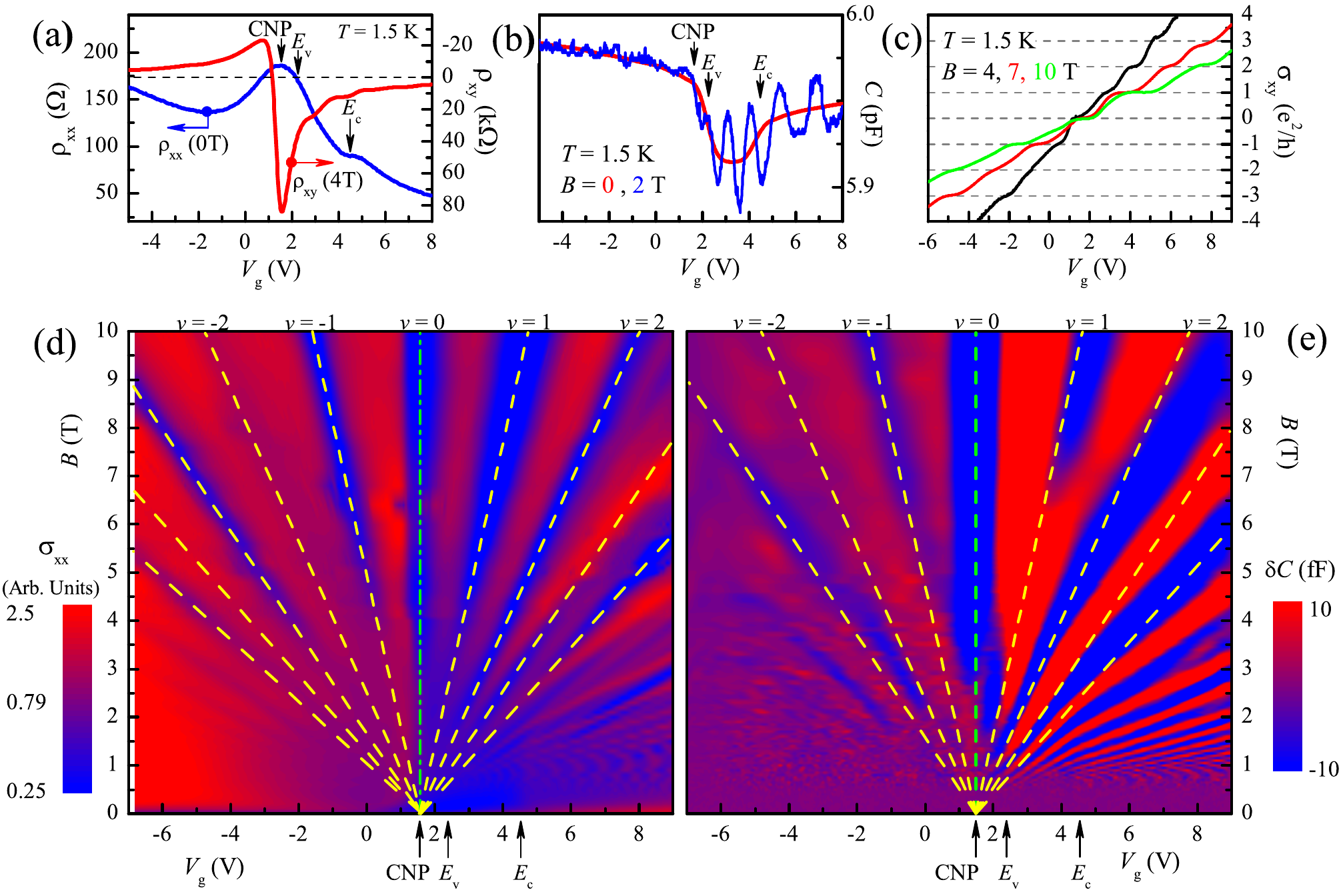}
\caption{\label{fig2} (a) Typical $\rho_{xx}(V_g)$ and $\rho_{xy}(V_g)$  traces measured at $B = 0 $ and $B = 4$\,T, respectively. $\rho_{xx}$ displays a maximum and $\rho_{xy}$ changes sign in the vicinity of the charge neutrality point (CNP). (b) Capacitance measured at $B= 0 $ and $B = 2$\,T. The pronounced minimum of the zero field capacitance corresponds to the reduced DoS when $E_F$ is in the gap. From that we can estimate the band edges to be at $E_v = 2.2$\,V and $E_c = 4.4$\,V. The quantum oscillations of the capacitance stem from the oscillations of the DoS. (c) Hall conductivity $\sigma_{xy}(V_g)$ measured for $B = 4$\,T (black), 7\,T (red), and 10\,T (green). Quantized steps occur both on the hole side and on the electron side. (d) 2D color map of $\sigma_{xx}(V_g, B)$ . For the plot each $\sigma_{xx}(V_g)$ trace  has been normalized with respect to the average value, set to 1. The red color stands for $\sigma_{xx}$ maxima while blue color displays minima. From the distance between blue $\sigma_{xx}$ minima we extract a filling rate of $\alpha_{\rm total} = 7.6\times 10^{10}$\,cm$^{-2}$/V (see text). This allows constructing a LL fan chart (dashed yellow lines) which describes the low filling factors $\nu$ well. (e) 2D color map of $\delta C(B) = C(B) - C(0)$ shown as function of $B$ and $V_g$. As in (d), the red color stands for LL maxima (DoS maxima) while blue color display the gaps between LLs. The LL fan chart is the same as in d. CNP, $E_c$, and $E_v$ are marked on the x-axis. }
\end{figure}

\twocolumngrid

Typical $\rho_{xx}$ and $\rho_{xy}$  traces as function of the gate voltage  $V_g$ are shown in Fig.~2a. $\rho_{xx}$ displays a maximum near $V_g = 1.5$\,V, whereas $\rho_{xy}$  changes sign; this occurs in the immediate vicinity of  the charge neutrality point (CNP)\cite{Kozlov14}. The corresponding capacitance $C(V_g)$ at $B = 0$ in Fig.~2b exhibits a broad minimum between 2.2\.V and 4.5\,V and echoes the reduced density of states $D_{\rm t}$ and $D_{\rm b}$  of the Dirac 2DES when the Fermi energy $E_F$ is in the gap of HgTe. For $V_g>4.5$\,V, $E_F$ moves into the conduction band where surface electrons coexist with the bulk ones. There, the capacitance (and thus the DoS) is increased and grows only weakly with increasing $V_g$. Reducing $V_g$ below 2.2\,V shifts $E_F$  below the valence band edge so that surface electrons and bulk holes coexist. A strong positive magnetoresistance, a non-linear Hall voltage and a strong temperature dependence of $\rho_{xx}$  provide independent confirmation that $E_F$  is in the valence band\cite{Kozlov14}. Due to the valley degeneracy of holes in HgTe and the higher effective mass, the DoS, and therefore the measured capacitance $C$ is highest in the valence band.

For  $B$ well below 1\,T both $C(V_g)$ and $\rho_{xx}(V_g)$ start to oscillate and herald the formation of Landau levels (LLs). The  $C(V_g)$ trace oscillates around the zero field capacitance, shown for $B=2$\,T in Fig.~2b. These oscillations, reflecting oscillations in the DoS, are more pronounced on the electron side (to the right of the CNP). This electron-hole asymmetry stems mainly from the larger hole mass, leading to reduced LL separation on the hole side. At higher fields Hall conductivity $\sigma_{xy}$ and resistivity $\rho_{xy}$ (not shown) become fully quantized, for both electron and hole side.  $\sigma_{xy}(V_g)$, shown for $B = 4$\,T, 7\,T, and 10\,T in Fig.~2c, shows quantized steps of height $e^2/h$ , ( $h$ = Planck's constant) as expected for spin-polarized 2DES.

An overview of transport and capacitance data in the whole $V_g$ and $B$ space is presented in Figs.~3d and 3e as 2D color maps (see the Supplementary Material for additional data). We start with discussing the $\sigma_{xx}$ data, calculated from $\sigma_{xx} = \rho_{xx}/(\rho_{xx}^2 + \rho_{xy}^2)$, in Fig.~2d first. The sequence of $\sigma_{xx}$ maxima and minima is almost symmetrical to the CNP where the electron and hole densities are equal. At fixed magnetic field the distance   $\Delta V_g$ between neighboring $\sigma_{xx}$ minima corresponds to a change in a carrier density $\Delta n$  from which we can calculate, with the LL degeneracy $eB/h$, the filling rate $dn/dV_g = \alpha_{\rm total} = 7.6\times10^{10}$\,cm$^{-2}$/V at 10\,T. Comparison of electron densities extracted in the classical Drude regime with densities taken from the periodicity of  SdH oscillations have shown that $\sigma_{xx}$  oscillations at high  $B$ reflect the total carrier density in the TI, i.e. charge carrier densities in the bulk plus in top and bottom surfaces \cite{Kozlov14}. We therefore conclude that the filling rate $\alpha_{\rm total}$ describes the change of the total carrier density $n$  with  $V_g$ and is directly proportional to the capacitance per area   $C/A = edn/dV_g$.  $\alpha_{\rm total}=  7.6\times10^{10}$\,cm$^{-2}$/V  corresponds to a capacitance of  $C/A = e \alpha_{\rm total} = 1.22\times10^{-4}$\,F/m$^{2}$, a value close to the calculated capacitance $C_{\rm gt}^{\rm calc}/A = 1.45\times 10^{-4}$\,F/m$^{2}$ using thickness and dielectric constant of the layers (see Figs.~1a and c and the Supplemental Material). Using this  $\alpha_{\rm total}$ extracted at 10\,T, the Landau level fan chart, i.e. the calculated positions of the $\sigma_{xx}$ minima as function of $V_g$ and $B$, fits the data for low filling factors $\nu$ quite well. For filling factors larger than $\nu = 2$  on the electron side the fan chart significantly deviates from experiment and is discussed using higher resolution data below. On the hole side where the SdH oscillations stem from bulk holes, fan chart and experimental data match almost over the whole $(V_g, B)$ range. Remarkably, minima corresponding to odd filling factors are more pronounced than even ones, reflecting the large hole $g$-factor.  We now turn to the magneto-capacitance data $\delta C = C(V_g, B) - C(V_g, B = 0)$ shown in Fig.~2e. The data are compared with the same fan chart derived from transport. On the electron side, the experimental $\delta C$ minima display a reduced slope compared to the calculated fan chart, pointing to a reduced filling rate. This is a first indication that the capacitance does not reflect the total carrier density in the system but predominantly the fraction of the top 2DES only. On the hole side the LL fan chart fits the data quite well but in contrast to transport, LL features are less well resolved there. This asymmetry is connected to the different effective masses; the enhanced visibility in transport is due to that fact that SdH oscillations depend on $D^2$ while the capacitance depends on  $D$ only.

\onecolumngrid

\begin{figure}[t]
\includegraphics[width=\columnwidth,keepaspectratio]{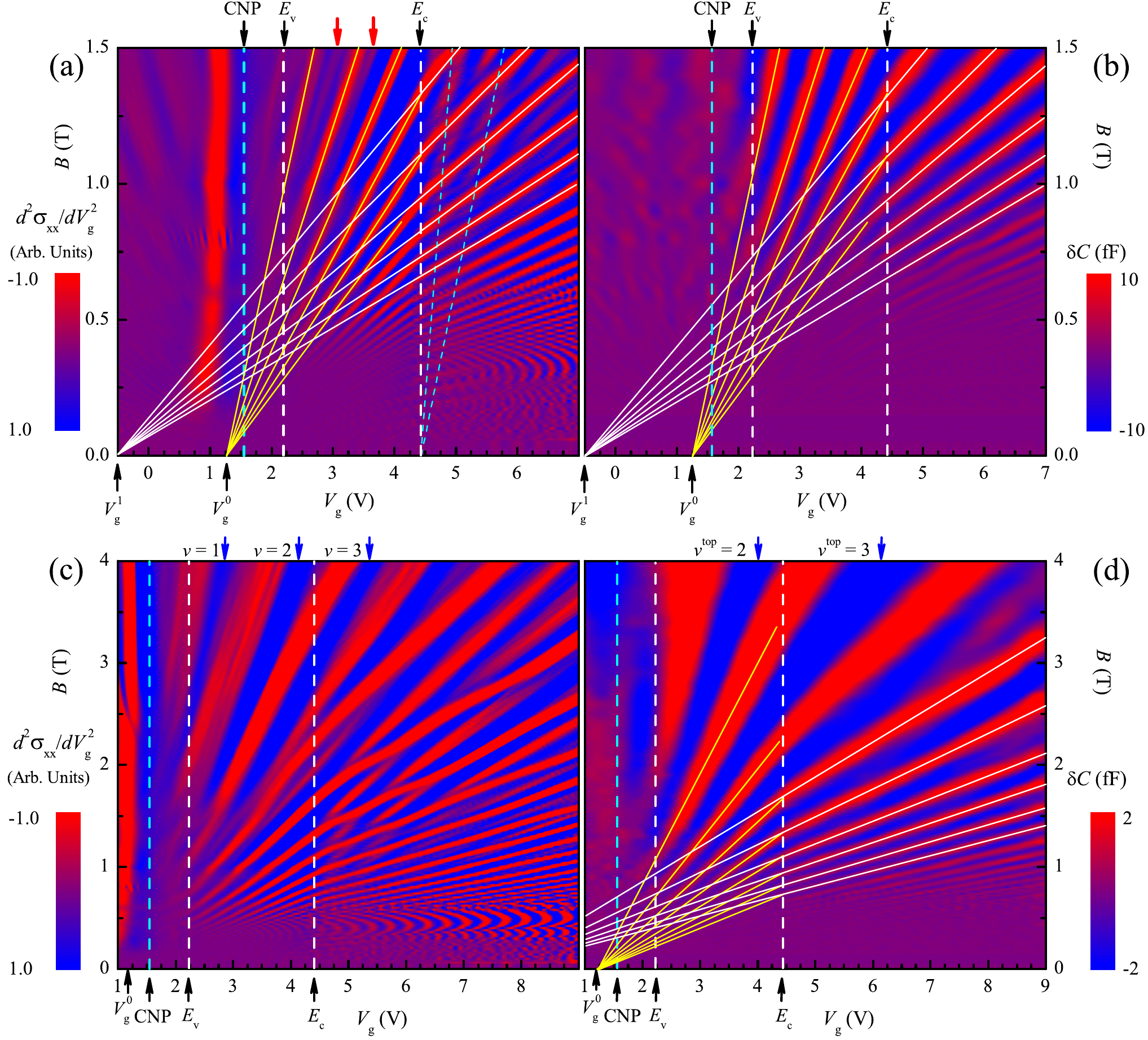}
\caption{\label{fig3} (a) 2D color map of $d^2 \sigma_{xx}/dV_g^2$ at low fields between 0 and 1.5\,T. The second derivative brings out quantum oscillation more clearly. In each row the root mean square value was normalized to 1. Red color again indicates large $\sigma_{xx}$ values, blue color low $\sigma_{xx}$, corresponding to LL gaps. The experimental data charts can be fitted by two fan charts originating at $V_g^0$ and $V_g^1$. The different slope of the two fan charts resembles the different filling rates $\alpha_{\rm top}^{\rm gap}$ and  $\alpha_{\rm top}^{\rm bulk}$ (see text). The red arrows and the two dashed lines starting at $V_g \approx 4.4$\,V mark extra features discussed in the text; the gap region and CNP are marked by dashed vertical lines. (b) Corresponding $\delta C(V_g, B)$ data. The two fan charts are the same as the ones in (a). (c) As (a), but for fields between 0 and 4\,T. For   $V_g > 4.5$\,V the data can no longer be described by simple fan charts; the pattern is entangled in a complicated way, suggesting that electrons of top and bottom surface and the bulk contribute to the conductivity oscillations. (d) Capacitance data corresponding to (c) show a quite regular Landau fan chart. The yellow and white lines in the gap and for $E_F < E_C$, respectively, correspond to the same filling rates $\alpha_{\rm top}^{\rm gap}$ and $\alpha_{\rm top}^{\rm bulk}$ as in (a) and (b). This indicates that the capacitance oscillations stem from the top surface only.}
\end{figure}

\twocolumngrid

Previous transport experiments have shown that the periodicity of the SdH oscillations is changed at low magnetic fields, reflecting a reduced carrier density. Tentatively, this was ascribed to SdH oscillations stemming from the top surface only \cite{Kozlov14}. When top and bottom surface electrons have different electron densities and mobilities it is expected that Landau level splitting - in sufficiently low $B$ - is in transport first resolved for the layer with the higher density and mobility, i.e. higher partial conductivity and lower LL level broadening. This expectation is consistent with previous experimental observation \cite{Kozlov14}. If, indeed, the low field SdH oscillations resemble the carrier density of a single Dirac surface, capacitance oscillations which probe preferentially the top surface and SdH oscillations should have the same period. Thus we compare $\sigma_{xx}$ and capacitance $\delta C$ in Fig.~3a and 3b at low $B$ up to 1.5\,T. In this low field region the capacitance (Fig.~3b) shows overall uniform oscillations of $\delta C$. The position of the maxima, corresponding to different LLs, is perfectly fitted by two fan charts featuring a distinct crossover at about $V_g = 4.4$\,V. The crossover is due to $E_F$ entering the conduction band which causes a reduced filling rate of the electrons responsible for the quantum oscillations at $V_g>4.4$\,V. Apparently, the quantum oscillations are caused by the top surface electrons only while the bulk electrons and electrons on the bottom surface merely act as a reservoir. Note, that the filling rate into the top surface state, probed in experiment, gets reduced when the filling rate into bulk states  $\neq 0$, since the total filling rate must be constant \cite{Kozlov14}. From the distance of $\delta C$  maxima at constant $B$ we can extract the filling rate $\alpha_{\rm top}^{\rm gap}$ in the gap (2.2\,V $<V_g <$ 4.4\,V) and when $E_F$  is in the conduction band, $\alpha_{\rm top}^{\rm bulk}$. From Fig.~3b we obtain  $\alpha_{\rm top}^{\rm gap}=5.25\times 10^{10}$\,cm$^{-2}$/V and $\alpha_{\rm top}^{\rm bulk}=3.3\times 10^{10}$\,cm$^{-2}$/V. This means that in the gap $\alpha_{\rm top}^{\rm gap}/\alpha_{\rm total}=70$\,\%
of the total filling rate apply to the top surface while the remaining 30\,\%
can be ascribed to the bottom surface. The reduced filling rate $\alpha_{\rm top}^{\rm bulk}$  for $E_F$  in the conduction band is $0.44\cdot \alpha_{\rm total}$ and hence the remaining filling rate of 56\,\%
is shared between bulk and back surface filling. We note that we obtain reasonable values for the filling rates only when we assume spin-resolved LL degeneracy. Since there is no signature of spin splitting down to 0.6\,T, where the oscillations get no longer resolved one can conclude that the quantum oscillations always stem from non-degenerate LLs, proving the topological nature of the charge carriers. The extrapolation of the two fan charts towards $B \rightarrow 0$  defines two specific points on the $V_g$-axis, denoted as $V_g^{0}=1.25$\,V  and $V_g^{1}=-0.5$\,V. These points correspond to vanishing electron density on the top surface $n_{\rm top}$ in case the respective filling rates $\alpha_{\rm top}^{\rm gap}$ and $\alpha_{\rm top}^{\rm bulk}$  would stay constant over the entire $V_g$ range. This is not the case as  $\alpha_{\rm top}^{\rm bulk} = {\rm constant}$ only applies for $E_F$  in the conduction band and $\alpha_{\rm top}^{\rm gap} = {\rm constant}$  for $E_F$ in the gap. Moving  $E_F$ into the valence band greatly reduces this filling rate. Therefore $V_g^{0}$ and  $V_g^{1}$ correspond only to virtual zeroes of the electron density while the real one is much deeper in the valence band.

To get a better resolution of the low field SdH oscillation we plot $d^2 \sigma_{xx} / dV_g^2$ in Fig.~3a; as before red regions indicate $\sigma_{xx}$  maxima. The same LL fan chart in Figs.~3b and~3a, fits both, transport and capacitance, quite well. Hence, at sufficiently low $B$, both $\sigma_{xx}$ and $\delta C$ oscillations resemble the carrier density of the top surface only. However, even at low $B$ striking deviations in $d^2 \sigma_{xx} / dV_g^2$ pop up which get more pronounced at higher $B$ (see below). Two faint lines at $B = 1 \ldots 1.5$\,T and $V_g = 2.5 \ldots 3.5$\,V which appear between the LLs of the top surface and are marked by red arrows in Fig.~3a, are ascribed to SdH oscillations stemming from the bottom surface. More differences occur once   $E_F$ enters the conduction band ($V_g > 4.4$\,V). New maxima appear in the fan chart and form rhomb like structures. These anomalous structures, completely absent in capacitance, get even more intriguing at higher $B$, displayed in Fig.~3c. Data taken at $B$ up to 4\,T, displayed in Figs.~3c and 3d, show marked differences between transport and capacitance. $\delta C$ maxima, given by the red regions in Fig.~3d are, as before, well described by the same two LL fan chart with the same filling rates as in Figs.~3a and b. The filling factors given on top of Fig.~3d are the ones of the electrons in the top surface only, while
the filling factors given in Fig.~3c are the ones of the total carrier density, determined by the total filling rate and the position of the $\sigma_{xy}$ plateaus in Fig.~2c. The transport data for $E_F$ in the gap show splitting of the Landau levels and for  $V_g > 4.4$\,V, i.e. for $E_F$ in the conduction band, a very complex structure with crossing LLs evolves, which is strikingly different from the one observed in the capacitance data. We thus conclude that in transport experiments the three available transport channels (top, bottom surface electrons, bulk electrons) contribute to the signal and lead to a complicated pattern of the quantum oscillations as a function of $B$ and $V_g$. The oscillations of the quantum capacitance, in contrast, stem preferentially from the top surface and allow probing the LL spectrum of a single Dirac surface in a wide range of $B$ and $V_g$. Corrections to that likely occur at high $B$ and $V_g$: a level splitting at  ($V_g \approx 7$\,V, $B\approx 3$\,T)  in Fig.~3d suggest that signals from bulk or back surface can affect also $\delta C$ at higher $B$, although to a far lesser degree when compared to transport.

In summary, we present first measurements of the quantum capacitance of a TI which directly reflect the DoS of Dirac surface states. The oscillations of the quantum capacitance in quantizing magnetic fields allow tracing the LL structure of a single Dirac surface. The complimentary information provided by transport and capacitance experiments is promising in getting a better understanding of the electronic structure of TIs, the latter being particularly important for potential applications of this new class of materials.

\acknowledgments We acknowledge funding by the Elite Network of Bavaria and by the German Science Foundation via SPP 1666. This work was partially supported by RFBR grants No 14-02-31631, 15-32-20828 and 15-52-16008.

\onecolumngrid

\newpage

\begin{center}
\Large{\textbf{Supplementary Information: ``Probing quantum
capacitance in a 3D topological insulator"}}

\large{D.\,A.~Kozlov, D.~Bauer, J. Ziegler, R.~Fischer, M.\,L.~Savchenko,
Z.\,D.~Kvon, N.\,N.~Mikhailov, S.\,A.~Dvoretsky and D.~Weiss}
\end{center}

\renewcommand{\thefigure}{S\arabic{figure}}

\section{Samples details}

\begin{figure}[h]
\includegraphics[width=0.95\linewidth]{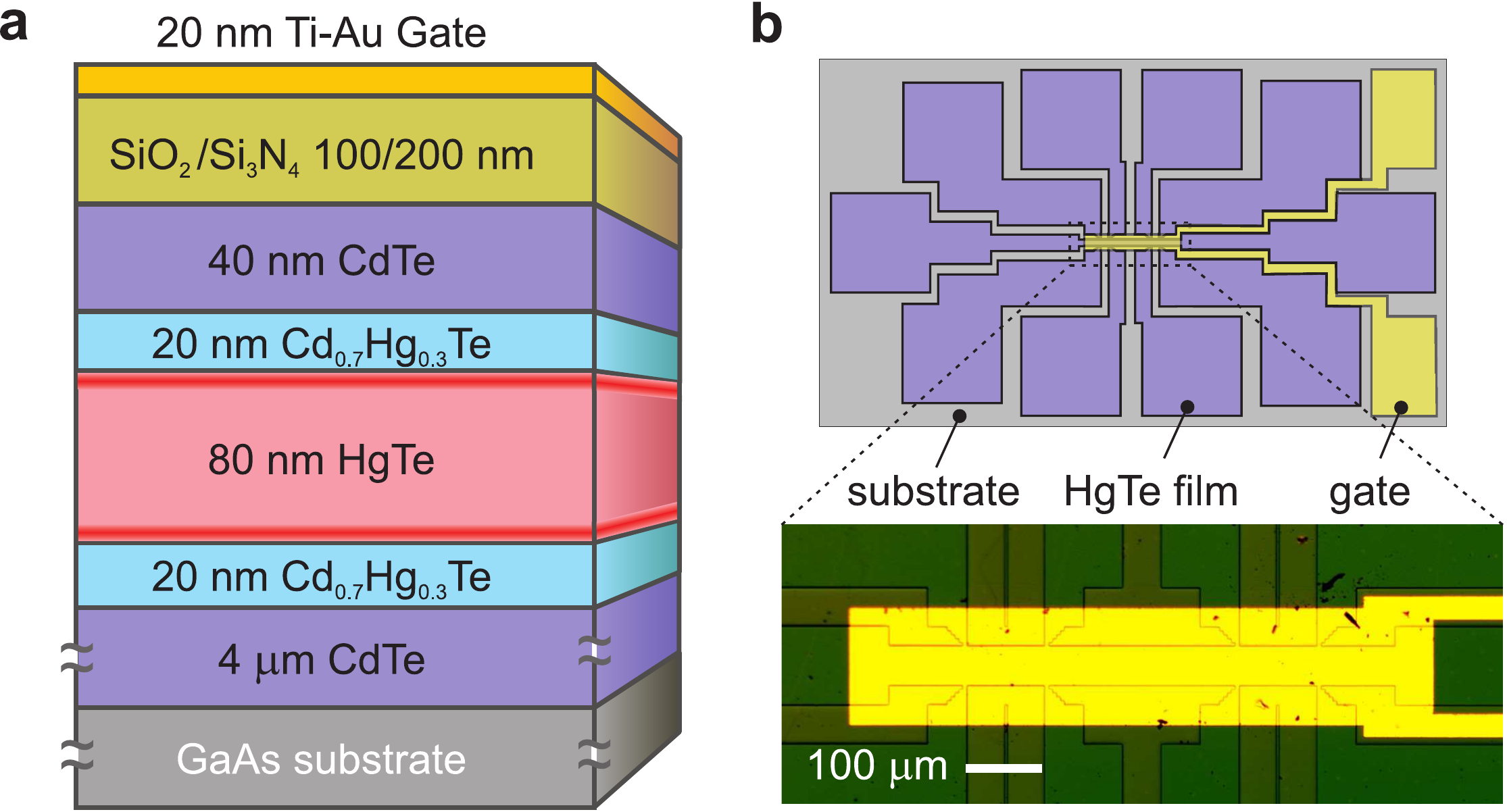}
\caption{\textbf{a}, Cross section of the heterostructure with SiO$_2$/Si$_3$N$_4$ insulator. The
Dirac surface states, shown in red, enclose the strained HgTe
layer. \textbf{b}, Top view of the device showing the Hall geometry with 10
potential probes, covered by the top gate (yellow); the gate is shown in more detail in the optical micrograph on the bottom.} \label{heterostructure}
\end{figure}

The experiments were carried out on strained 80\,nm thick HgTe films,
grown by molecular beam epitaxy (MBE) on CdTe (013)
\cite{DantscherPRB14}. The Dirac surface electrons have high
electron mobilities of order  $\mu\approx
4\times10^5$\,ñm$^2$/V$\cdot$s \cite{Kozlov14}. The cross
section of a structure is sketched in
Fig.~\ref{heterostructure}a. For transport and capacitance
measurements, carried out on one and the same device, the films
were patterned into Hall bars with metallic top gates
(Fig.~\ref{heterostructure}b).

For gating, two types of dielectric layers were used, giving
similar results: 100\,nm SiO$_2$ and 200\,nm of Si$_3$N$_4$ grown
by chemical vapor deposition or 80\,nm Al$_2$O$_3$ grown by atomic
layer deposition (not shown in Fig.~\ref{heterostructure}a). In
both cases, TiAu was deposited as a metallic gate. The
measurements of the total capacitance $C$ between the gate and the
HgTe layer were performed at temperature $T=1.5$\,K and in
magnetic fields up to 13\,T. Several devices from the same wafer
have been studied. For magnetotransport measurements standard
lock-in technique has been applied with excitation AC currents of
10-100\,nA at frequencies $\omega/2\pi$ ranging from 0.5 to
12\,Hz. For the capacitance measurements we superimpose the DC
bias voltage with a small AC bias voltage (see Fig. 2c) and
measure with lock-in technique the AC current flowing across our
device phase sensitively. The typical AC voltage was 50\,mV at a
frequency of 213\,Hz. The absence of both leakage currents and
resistive effects in the capacitance were controlled by the real
part of the measured AC current. These resistive effects occur
when the condition $RC \omega <<1$ becomes invalid, for example
when the series resistance $R$, i.e. the resistance of the two-dimensional electron gas
increases at integer filling factors and high magnetic fields. In
order to suppress resistive effects in high magnetic fields
additional measurements at lower frequencies (down to several Hz)
were performed (see below).

\section{Samples band diagram and equivalent circuit}
\subsection{Electrostatic and band diagram}

Following the papers of Stern \cite{Stern83}, Smith \cite{Smith85}
and Luryi \cite{Luryi88} we introduce an equivalent circuit for the
sample's capacitance based on simple one-dimensional electrostatics. The
discussion is limited to the  case where the Fermi level is
located in the gap, so that contributions of bulk holes and
electrons are absent. The top and bottom surface layers are
treated as infinitely thin, negatively charged surfaces with
finite density of states (DoS), i.e. we ignored the spatial
distribution of the surface electrons. Finally, we treated the
insulating layer as uniform with an average dielectric constant
and with the total thickness of the layers given in
Fig.~\ref{heterostructure}.

\begin{figure}[t]
\includegraphics[width=0.95\linewidth]{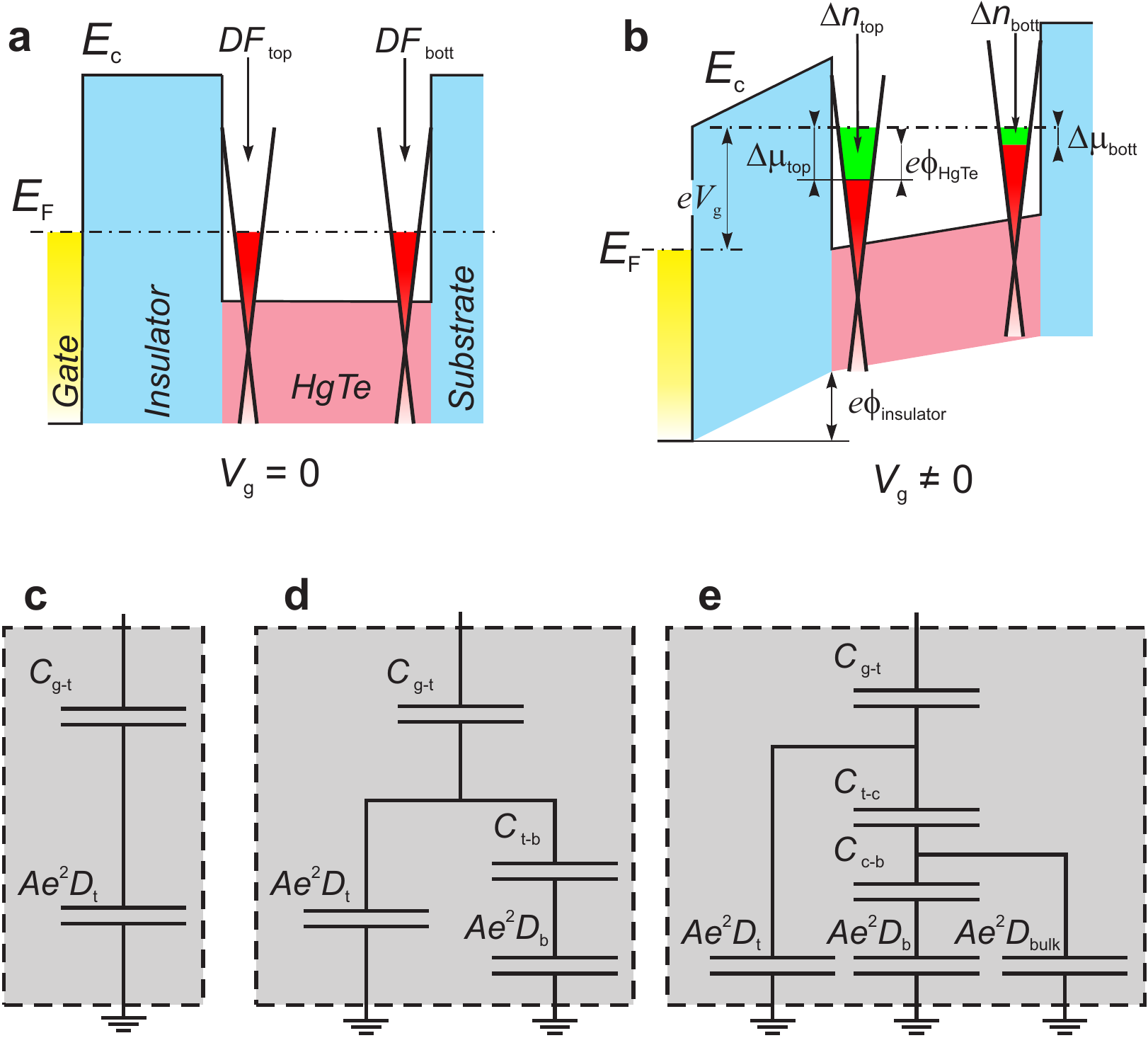}
\caption{\textbf{a} and \textbf{b}, Simplified band diagrams of the structure
studied for $E_{\rm F}$ is in the bulk gap at $V_g = 0$
(panel a) and $V_g>0$ (panel b). \textbf{c}, \textbf{d} and \textbf{e}, Equivalent
circuit for a simple capacitor where one plate consists of a two-dimensional electron gas (c), for our three plate capacitor with $E_{\rm F}$ in the bulk gap (d) and for a three plate capacitor with $E_{\rm F}$ in the conduction band or valence band, respectively (e).} \label{Band_diagram}
\end{figure}

A cartoon of the resulting simplified band diagram  is presented
for zero gate voltage $V_g$ in Fig.~\ref{Band_diagram}a. Without
applied $V_g$ the Fermi level is constant across the structure.
The density of  electrons $n_{\rm t}$ and $n_{\rm b}$ on  top and
bottom surface, respectively, depends on the position of the Fermi
level with respect to the Dirac point (which is, other than
sketched in Fig.~\ref{Band_diagram}a below the valence band edge).
For the sake of simplicity we assume that the densities are the
same at $V_g = 0$. This  simplification is justified since we are
only interested in changes of the carrier density with gate
voltage and not in absolute values of the carrier density.

For an applied gate voltage $V_g \neq 0 $ we assume that the
voltage drop occurs exclusively across the SiO$_2$/Si$_3$N$_4$
insulator leading to an energy difference $e V_g$ between the
Fermi level in the gate metal and the HgTe layer. Here, $e$ is the
elementary charge. Applying a positive gate voltage $V_g >0$
increases the carrier density in top and bottom layer by $\Delta
n_{\rm top}$ and $\Delta n_{\rm bott}$, respectively. Due to
screening of the electric field by the top surface layer the
induced electron density is higher in the top layer than in the
bottom one; while the Fermi level throughout the HgTe layer is
constant the different electron density in top and bottom
surface layer causes an electrical potential drop between bottom
and top surfaces of $e\phi_{\rm HgTe}$. The change of carrier
density in top and bottom layer can be written as $\Delta \mu_{\rm
top (b) }= \Delta n_{\rm top (bott)} / D_{\rm t (b)}$, where
$D_{\rm t(b)}$ is the density of states (DoS) of the electrons on
the top (bottom) surface. Thus the following relation holds:

\begin{equation} \label{form1}
    \Delta \mu_{\rm top} = \Delta \mu_{\rm bot} + e \phi_{\rm
    HgTe}.
\end{equation}

Using charge neutrality and Gauss's law of electrostatics the potential drop in the HgTe layer can be calculated from the electric field  $E_{\rm HgTe} = e \Delta n_{\rm b} /
\varepsilon_{\rm HgTe} \varepsilon_0$, induced by the additional
electron density  $\Delta n_{\rm b}$ on the bottom layer:
\begin{equation} \label{form2}
    \phi_{\rm HgTe} = E_{\rm HgTe} d_{\rm HgTe} =
    e \Delta n_{\rm bot} d_{\rm
    HgTe} / \varepsilon_{\rm HgTe} \varepsilon_0 .
\end{equation}
$\varepsilon_{\rm HgTe}$ is the dielectric constant of HgTe and
$d_{\rm HgTe}$ is the distance between the electron wave functions of top and bottom surfaces layers, approximated by the thickness of the HgTe layer, $d_{\rm HgTe}$.  Combining formulas \ref{form1} and
\ref{form2} gives a relation between $\Delta n_{\rm
top}$ and $\Delta n_{\rm bot}$:
\begin{equation}
    \Delta n_{\rm top} / D _{\rm t}= \Delta n_{\rm bot} / D _{\rm b} +  e^2\Delta n_{\rm bot} d_{\rm
    HgTe} / \varepsilon_{\rm HgTe} \varepsilon_0 .
\end{equation}

This is the expression given by Luryi \cite{Luryi88} generalized
for non-parabolic dispersion of the surface electrons.

\subsection{The equivalent circuit}

In the main text we discuss the equivalent circuit of a three plate
capacitor as a model for the capacitance measured in a topological
insulator with metallic gate and top and bottom surface layers.
The equivalent circuit is applicable when the Fermi level is in
the bulk gap of HgTe and is shown in Fig.~1d and the Supplementary
Figure~\ref{Band_diagram}d. A similar equivalent circuit has been
recently suggested for the same system but a slightly different
configuration \cite{Baum14}. The four capacitors drawn in
Supplementary Figure~\ref{Band_diagram}d represent either
geometrical ($C_{gt}$ and $C_{tb}$) or quantum ($A e^2 D_{\rm t}$
and $A e^2 D_{\rm b}$) capacitances. Here, $C_{gt} = A
\varepsilon_0 \varepsilon_{\rm ins} / d_{\rm ins}$ is the geometrical
capacitance associated with the insulating layer between the gate
and the top surface of the HgTe layer; $C_{tb} = A \varepsilon_0
\varepsilon_{\rm HgTe} / d_{\rm HgTe}$ is responsible for the
potential drop across  the HgTe layer between electrons on the top
and bottom surface; $A e^2 D_{\rm t}$ and $A e^2 D_{\rm b}$ are
the corresponding quantum capacitances representing the finite DoS
of the corresponding HgTe surface states; in all cases $A$ is the
gate area. The equivalent circuit in Fig.~\ref{Band_diagram}d
describes the electrostatics of the system: Indeed, the charge on
the quantum capacitors is the one induced by the gate voltage,
$\Delta n_{\rm top}$ and $\Delta n_{\rm bot}$ and the voltage
across the geometrical ones is, as it should be, proportional to
the corresponding induced charge and the thickness of the
respective dielectric layer.

The circuit becomes simpler if one removes the influence of the
bottom surface: this can be done by increasing the distance
between top and bottom layer ($C_{\rm tb} \rightarrow 0$) or by
removing the charge carriers ($e^2 D_{\rm b}\rightarrow 0$). Then  only two capacitors are present in the reduced circuit
(Fig.~\ref{Band_diagram}c); this is the situation relevant for a
gated two-dimensional electron system. In contrast,  if the Fermi
level is lying in the bulk electron or hole band (see
Fig.~\ref{Band_diagram}e) the situation becomes more complex. In
this case bulk carriers are characterized by a particular wave
function across the HgTe layer. A simple approach to model this is
to replace the capacitor  $C_{\rm tb}$ by two capacitors $C_{\rm
tc}$ and $C_{\rm cb}$ representing the potential drop between  the
maximum of the bulk carriers' wave function (denoted by subscript
"c" at a point around the center of the HgTe layer) and the top and bottom of the HgTe layer, respectively. For
this the relation $C_{\rm tb}^{-1}=C_{\rm tc}^{-1} + C_{\rm
cb}^{-1}$ holds.

The total sample's capacitance $C$ for the circuit shown in
Fig.~\ref{Band_diagram}c can be easily derived using Ohm's law and is given by:

\begin{equation}\label{CSample}
\frac{A}{C}
    =  
    \frac{A}{C_{\rm gt}} + \frac{1}{e^2 D_{\rm t}
+ \frac{1}{ \frac{A}{C_{\rm tb}} + \frac{1}{e^2 D_{\rm b}}}}  =\\
           A C_{\rm gt}^{-1} +
        \Bigl[
            e^2 D_{\rm t} + \Bigl( A C_{\rm tb}^{-1} + (e^2 D_{\rm b})^{-1} \Bigr)^{-1}
        \Bigr]^{-1}
\end{equation}

In the actual device the relation $C_{\rm gt}/A \ll e^2 D_{\rm t},
e^2 D_{\rm b}$ holds, therefore $C \approx C_{\rm gt}$
and any variation of the DoS, e.g. in a magnetic field, leads only to a small change of the total capacitance $C$. An important result which can be derived from Eq.\, \ref{CSample}
is the sensitivity of the total capacitance  $C$  to a change
of the DoS of top and bottom layer, $D_{\rm t}$ and $D_{\rm b}$, respectively:

\begin{equation}
dC/dD_{\rm t}=
    \frac{
            A e^2 C_{\rm gt}^2 C_{\rm tb}^2
            ( C_{\rm tb} + A e^2 D_{\rm b} )^2
        }
        {
            \Bigl[
            C_{\rm gt}(C_{\rm tb}+A e^2 D_{\rm b} )+
            A e^2 \bigl( C_{\rm tb} ( D_{\rm b} + D_{\rm t} ) + A e^2 D_{\rm b} D_{\rm t} \bigr)
            \Bigr]^{2}
        }
\end{equation}

\begin{equation}
dC/dD_{\rm b}=
    \frac{
            A e^2 C_{\rm gt}^2 C_{\rm tb}^2
        }
        {
            \Bigl[
            C_{\rm gt}(C_{\rm tb}+A e^2 D_{\rm b} )+
            A e^2 \bigl( C_{\rm tb} ( D_{\rm b} + D_{\rm t} ) + A e^2 D_{\rm b} D_{\rm t}  \bigr)
            \Bigr]^{2}
        }
\end{equation}

The ratio of both quantities
\begin{equation}
\frac{
        dC/dD_{\rm t}
    }
    {
        dC/dD_{\rm b}
    }=
\frac{
        (C_{\rm tb} + A D_{\rm b} e^2)^2
    }
    {
        C_{\rm tb}^2
    }=
\Bigl(1 +
    \frac{
        A D_{\rm b} e^2 }
        { C_{\rm tb}}
\Bigr)^2
\end{equation}

is always larger than 1 as long as $D_{\rm b}\neq 0$  which means that the total capacitance $C$ is more sensitive to oscillations of the top layer, e.g.,  the same oscillation amplitude of $D_{\rm
t}$ and $D_{\rm b}$ will lead to  significantly different
oscillation amplitudes in  the measured total capacitance $C$.

\subsection{Numeric evaluations}

Here, we provide the values of the capacitances of our devices.
The insulating layer underneath the metallic gate consists of
Si$_3$N$_4$ (200\,nm layer with $\varepsilon = 7.5$), SiO$_2$
(100\,nm with $\varepsilon = 3.5$), CdTe (40\,nm with $\varepsilon
= 10.2$), CdHgTe (20\,nm with $\varepsilon \approx 13$) and HgTe
(5-7\,nm with $\varepsilon \approx 21$) and $C_{\rm gt}^{\rm calc}
/ A = 1.45 \times 10^{-10}$\,F/mm$^2$. The calculated value of the
specific geometrical capacitance $C_{\rm gt}^{\rm calc}/A$ is very
close to the experimental one, $C_{\rm gt}/A = e \alpha_{\rm total}
= 1.22\times 10^{-10}$\,F/mm$^2$ obtained from the filling rate
$\alpha_{\rm total} = 7.6\times10^{10}$\,cm$^{-2}$/V$\cdot$s (see
 main text).

Using the gate area $A = 0.05$\,mm$^2$ (see
Fig.~\ref{heterostructure}b) the sample's geometric capacitance is expected to
be $C_{\rm gt} \approx 6$\,pF which is in line
with the measured value ($\approx 23$\,pF) if one subtracts the
parasitic capacitance of our set up, being  around
17\,pF.

With the known values of the surface electrons' effective mass $m^* =
0.030 m_0$ \cite{DantscherPRB14} one can estimate the specific
quantum capacitances: $e^2 D_{\rm t(b)}^{\rm eval} = e^2 m^* /2
\pi \hbar^2 \approx 10^{-8}$\,F/mm$^2$ which is almost two orders of magnitude
higher than the specific geometrical capacitance $C_{\rm gt}^{\rm
calc}/A$.

Finally, the evaluated specific capacitance of the $80$\,nm thick HgTe layer with
$\varepsilon = 21$ is $C_{\rm tb}^{\rm calc}/A =
2.4\times 10^{-9}$\,F/mm$^2$. From this value and with Eq.\,3 one
can derive the relation between filling rates of electrons on  top
and bottom surfaces: $\alpha_{\rm top}/\alpha_{\rm
bott}\equiv(dn_{\rm top}/dV_g) / (dn_{\rm bot}/dV_g) = \Bigl(
\frac{D_{\rm t}}{ D_{\rm b}} + \frac{Ae^2 D_{\rm t} }{C_{\rm
tb}^{\rm eval}}  \Bigr) \approx 5.1$.
%
However, in our experiment, described in the main text, we
obtain $\alpha_{\rm top} = 0.7 \times \alpha_{\rm total}$,
$\alpha_{\rm bot} = 0.3 \times \alpha_{\rm total}$ and
$\alpha_{\rm top} / \alpha_{\rm bot} \approx 2.3$. This
discrepancy suggests that the equivalent circuit in
Fig.~\ref{Band_diagram}d does not fully describe the experimental
situation. Most likely this deviation stems from the fact that we
neglected the  dependence of the surface states' dispersion on the transversal electric field described
in \cite{Brune14}. Phenomenologically, this shortcoming can be
corrected by introducing a larger effective capacitance of $C_{\rm
tb}^{\rm calc}/A = 7.5\times 10^{-9}$\,F/mm$^2$, indicating an
enhanced screening in HgTe. This can be achieved by a higher value
of the dielectric constant of HgTe or a reduced effective layer
thickness.

We note that for both values of $C_{\rm tb}$ given above, the
ratio  $(dC/dD_{\rm t}) / (dC/dD_{\rm b})$ is larger than 5.5,
meaning that the sample capacitance is predominantly sensitive to
changes of the DoS of the top surface.

\newpage

\section{Densities and partial filling rates}

\begin{figure}[h]
\includegraphics[width=0.55\linewidth]{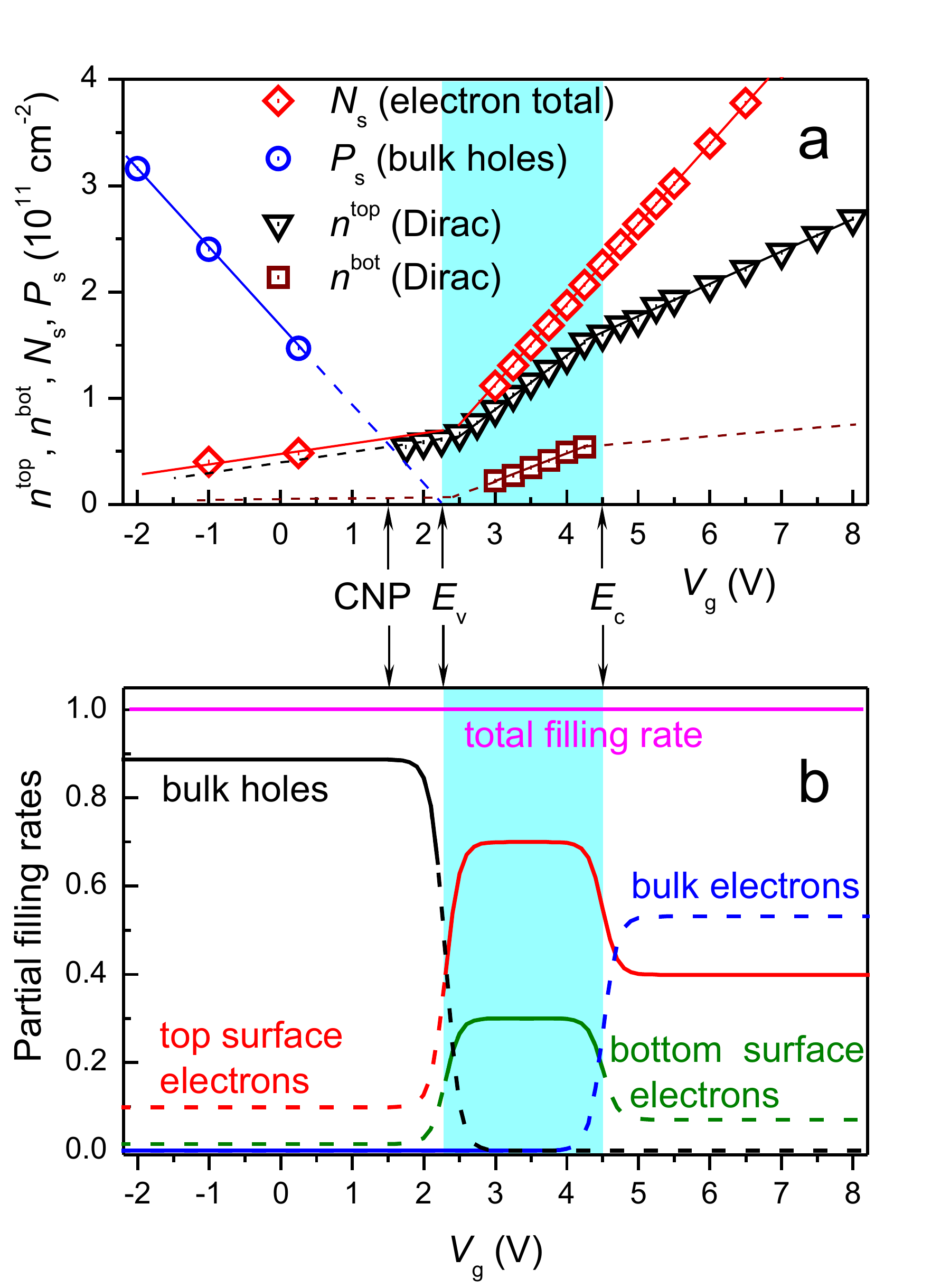}
\caption{\textbf{a}, Gate voltage dependence of the different charge carrier species. $n_{\rm top}$ is the electron density of the top surface; $n_{\rm bot}$ the one of the bottom surface; $N_{\rm s}$ is the total density of both surfaces plus the bulk electron density; $P_{\rm s}$ is the bulk hole density. The dashed lines extrapolate the corresponding
charge carrier species in regions where data can not be directly extracted. \textbf{b}, The absolute value of the partial filling rates  between surface and bulk carriers. The bluish area on both panels corresponds the bulk gap region. "CNP", "$E_v$" and "$E_c$" mark the charge neutrality point, top of the valence band and bottom of the conductive band, respectively. Values of the partial filling rates given by dashed lines can not be directly extracted from experiment but are taken from the corresponding equivalent circuit. }\label{Density}
\end{figure}

The gate voltage dependence of the different carrier species is shown in
Fig.~\ref{Density}a. The total electron density $N_{\rm s}$ and
the bulk hole density $P_{\rm s}$ were obtained from the low field transport data using the classical
(two-carrier) Drude model as described in \cite{Kozlov14}. The
top surface electron density $n_{\rm top}$  was obtained from the periodicity $\Delta (1/B)$ of the capacitance oscillations, $n_{\rm top}=e(h\Delta (1/B))^{-1}$, with the period $\Delta (1/B)$ on the $1/B$ scale and Planck's constant $h$.
 For $V_g < 1.5$\, this technique
is not applicable since the oscillations are dominated by bulk holes. The
density of the bottom surface electrons was obtained from
$n_{\rm bot } = N_{\rm s} - n_{\rm top}$ which holds as long as $E_{\rm
F}$ is located in the bulk gap, i.e. $V_{\rm g} =
2.2\ldots4.5$\,V. In our experiment it is impossible to directly
measure the density of electrons on the bottom surface since their
response is masked by electrons on the top surface. Consequently
the value of $n^{\rm bot }$ can not be derived when the Fermi
level is in the valence (because of unknown value of
$n^{\rm top}$) or the conductance band (because of the unknown value of
$n^{\rm bulk}$). However, based on the equivalent circuit shown in
Fig.~\ref{Band_diagram}d and e one can make an educated guess on the partial  filing rates. The result of such analysis is shown
in  Fig.~\ref{Density}b.

\newpage
\section{Resistive effects in the capacitance}
\begin{figure}[h]
\includegraphics[width=0.6\linewidth]{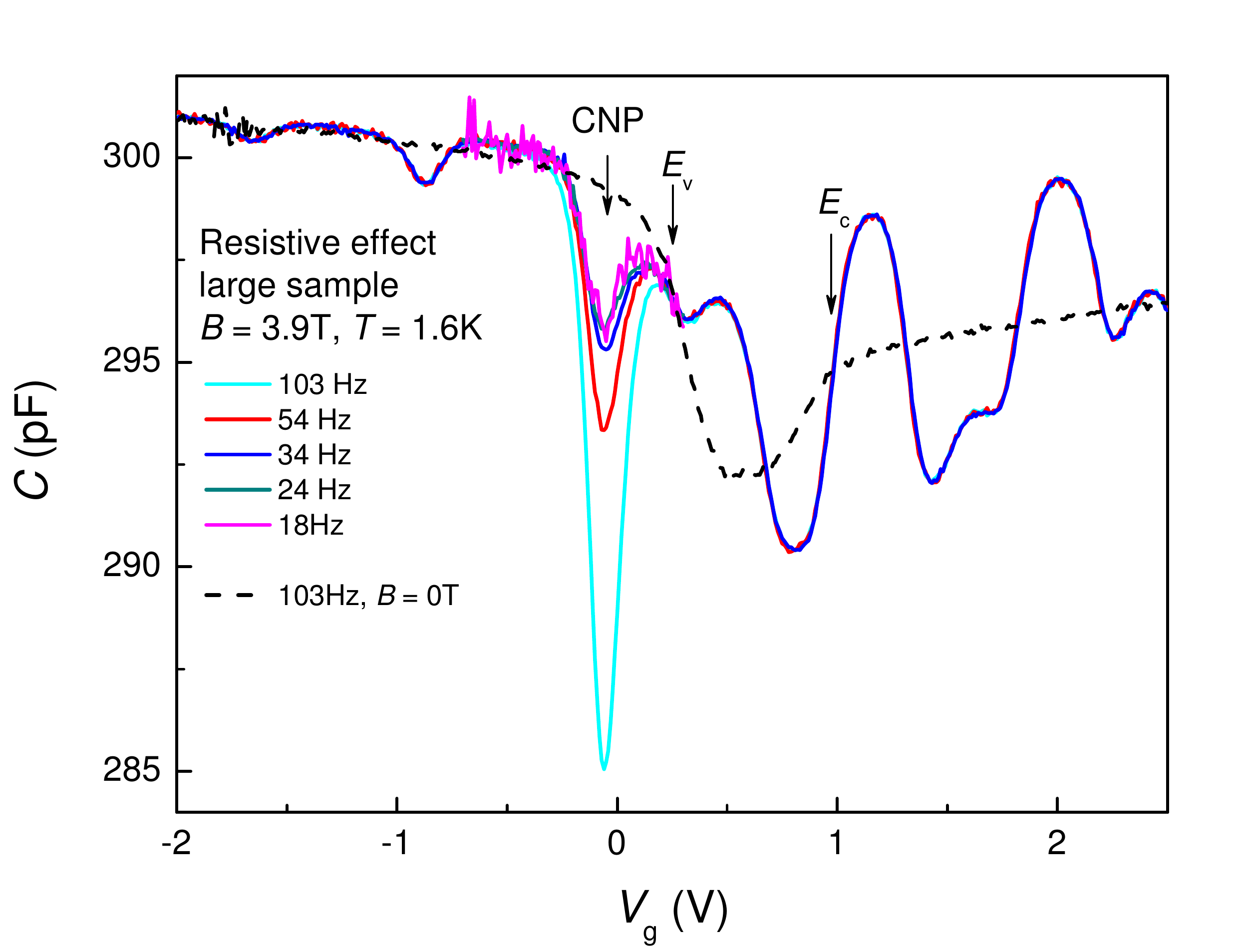}
\caption{Frequency dependence of the magnetocapacitance data:
$C(V_{\rm g})$  measured at a magnetic field of 3.9\,T and at
frequencies $f = 18, 24, 34, 54$ and $103$\,Hz. The dashed line
corresponds to the zero field $C(V_{\rm g})$ trace.}\label{ResistiveEffect}
\end{figure}

As mentioned in the main text,  resistive effects occur when the
condition $RC \omega <<1$ becomes invalid, for example when the
resistance $R$ of a two-dimensional electron gas connected in series to the capacitance increases
at integer filling factors and high magnetic fields. The absence
of both leakage currents and resistive effects in the capacitance
were controlled by the real part of the measured AC current. A direct manifestation of  resistive effects is the
frequency-dependence of the measured $C(V_{\rm g})$ traces. A corresponding  example, taken from a sample with larger gate area and therefore larger capacitance is shown in  Fig.~\ref{ResistiveEffect}. The capacitance minimum appearing at the charge neutrality point (CNP) and connected to a gap opening at higher magnetic fields shows a pronounced frequency dependence. The resistive effects are weak up to 34\,Hz. With increasing frequency, however, the minimum becomes deeper and the capacitance signal no longer reflects the density of states. Remarkably,  in all our samples
the resistive effects first appear close the CNP and are much less
pronounced at integer filling factors at the same values
of magnetic field. Thus the remaining capacitance oscillations reflect the density of states.

\newpage
\section{Experimental Data}

\begin{figure}[p]
\includegraphics[width=\linewidth]{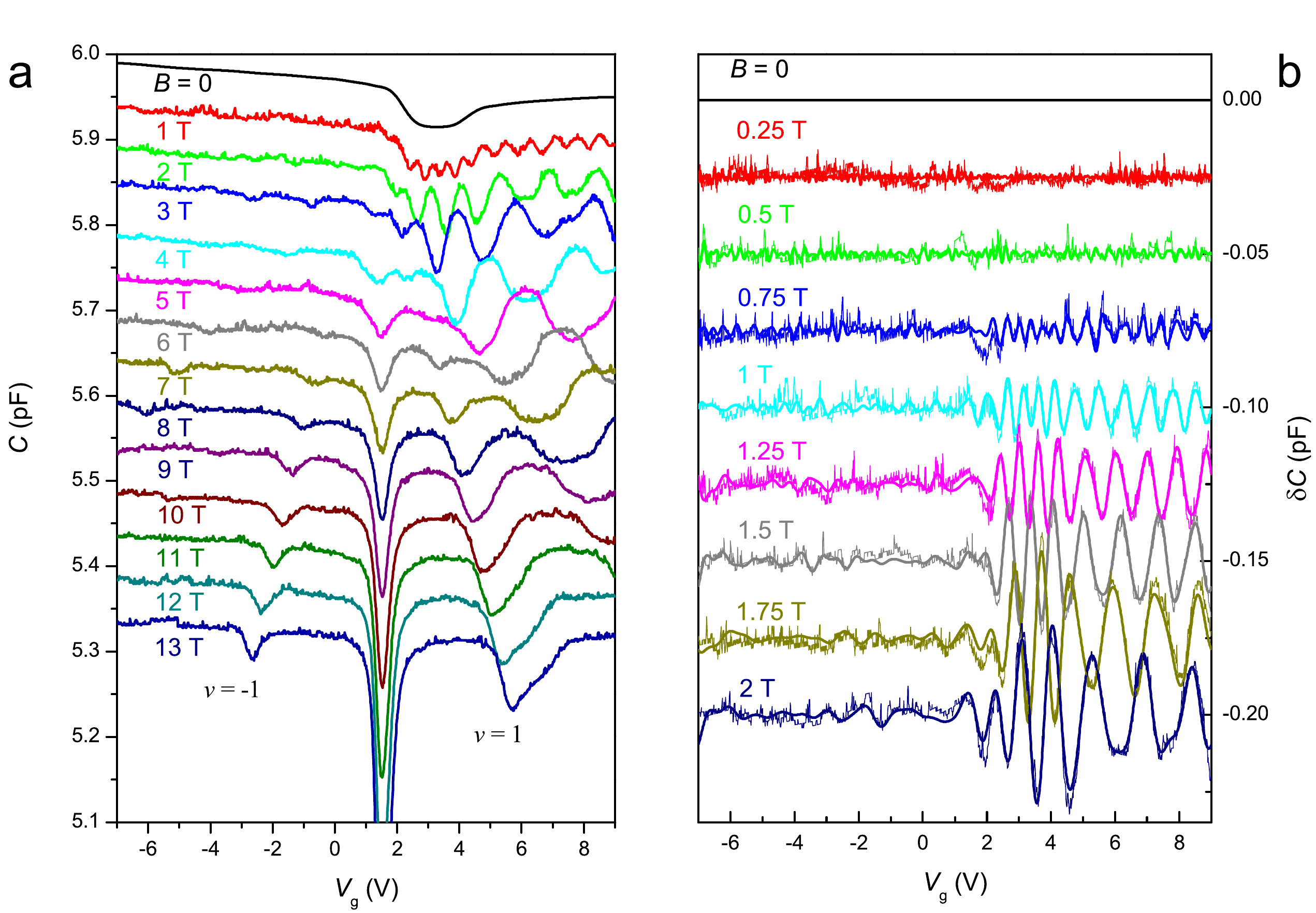}
\caption{Magnetocapacitance data: \textbf{a}, $C(V_g)$ dependencies at
magnetic fields from 0 (on top) to 13\,T (bottom) using
steps of 1\,T. Starting at $B=1$\,T each subsequent trace is shifted by -0.05\,pF on
the Y-axis.  The filling factors $v = -1$ and 1 are indicated. \textbf{b},
Differential magnetocapacitance $\delta C(V_g) =
C(V_g)|_{B}-C(V_g)|_{B=0}$ dependencies at magnetic fields from 0
(top) to 2\,T (bottom) using steps of 0.25\,T between the traces. Starting at $B=0.25$\,T each subsequent trace is shifted  by
-0.025\,pF on the Y-axis.  The thin lines
correspond to the raw data and the thicker ones are the data after
Fourier filtering. }\label{Cap_Vg}
\vfill
\end{figure}

\begin{figure}[p]
\includegraphics[width=\linewidth]{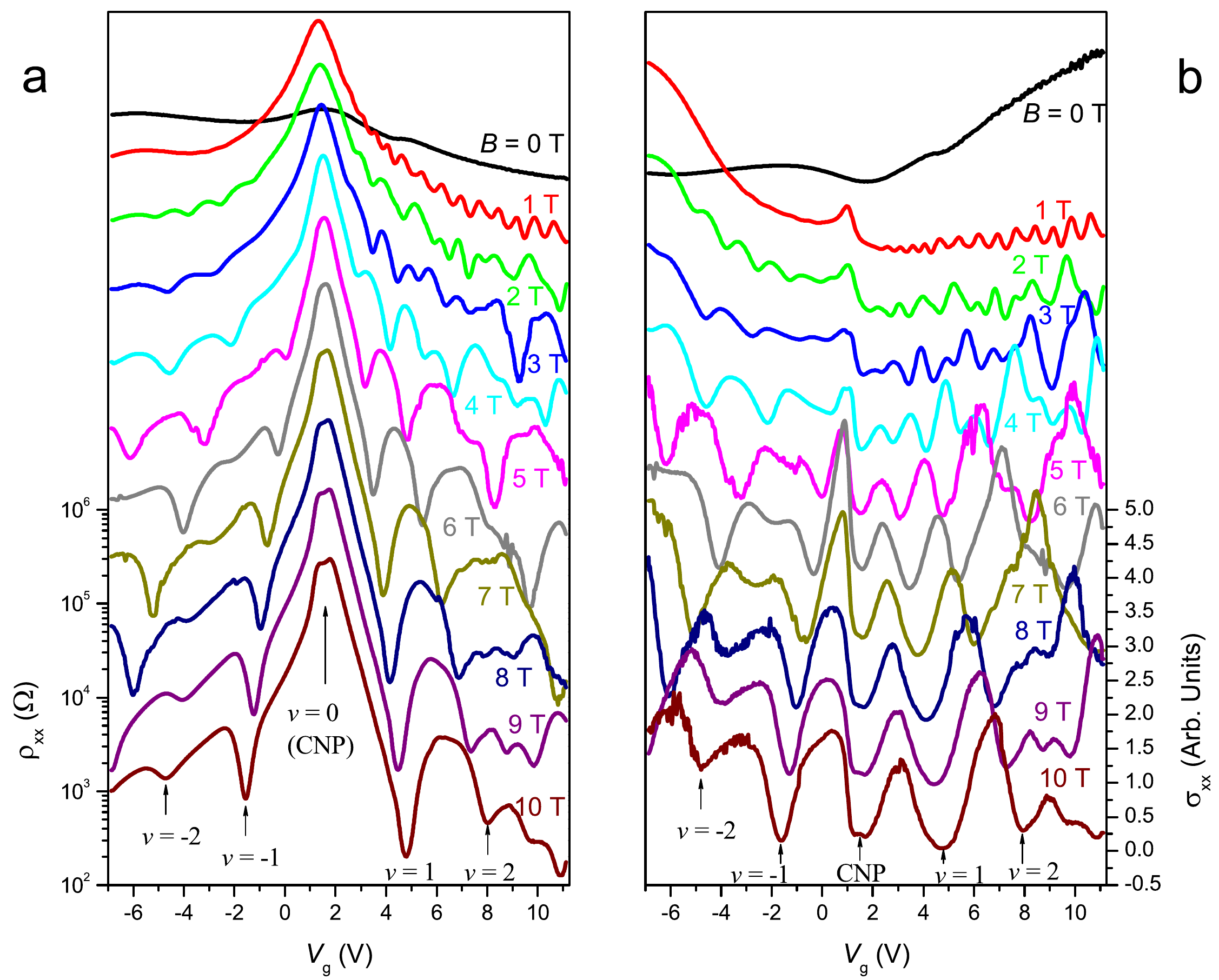}
\caption{Strong magnetic field magnetotransport data: \textbf{a},
$\rho_{xx}(V_g)$ dependencies at magnetic fields from 0 (top) to 10\,T (bottom).  Starting at
$B=9$\,T each subsequent trace is shifted by 1\,T on the Y-axis.  \textbf{b}, $\sigma_{xx}(V_g)$ dependencies from 0 (on
the top) to 10\,T (bottom).  Each trace was
normalized by the average value $<\sigma_{xx}(V_g)> =
1$ (see  text) and shifted by 1\,T on the Y-axis. The filling factors $\nu =$ -2, -1, 1, and 2 are marked.
}\label{Transport_StrongFields}
\end{figure}

\begin{figure}[p]
\includegraphics[width=\linewidth]{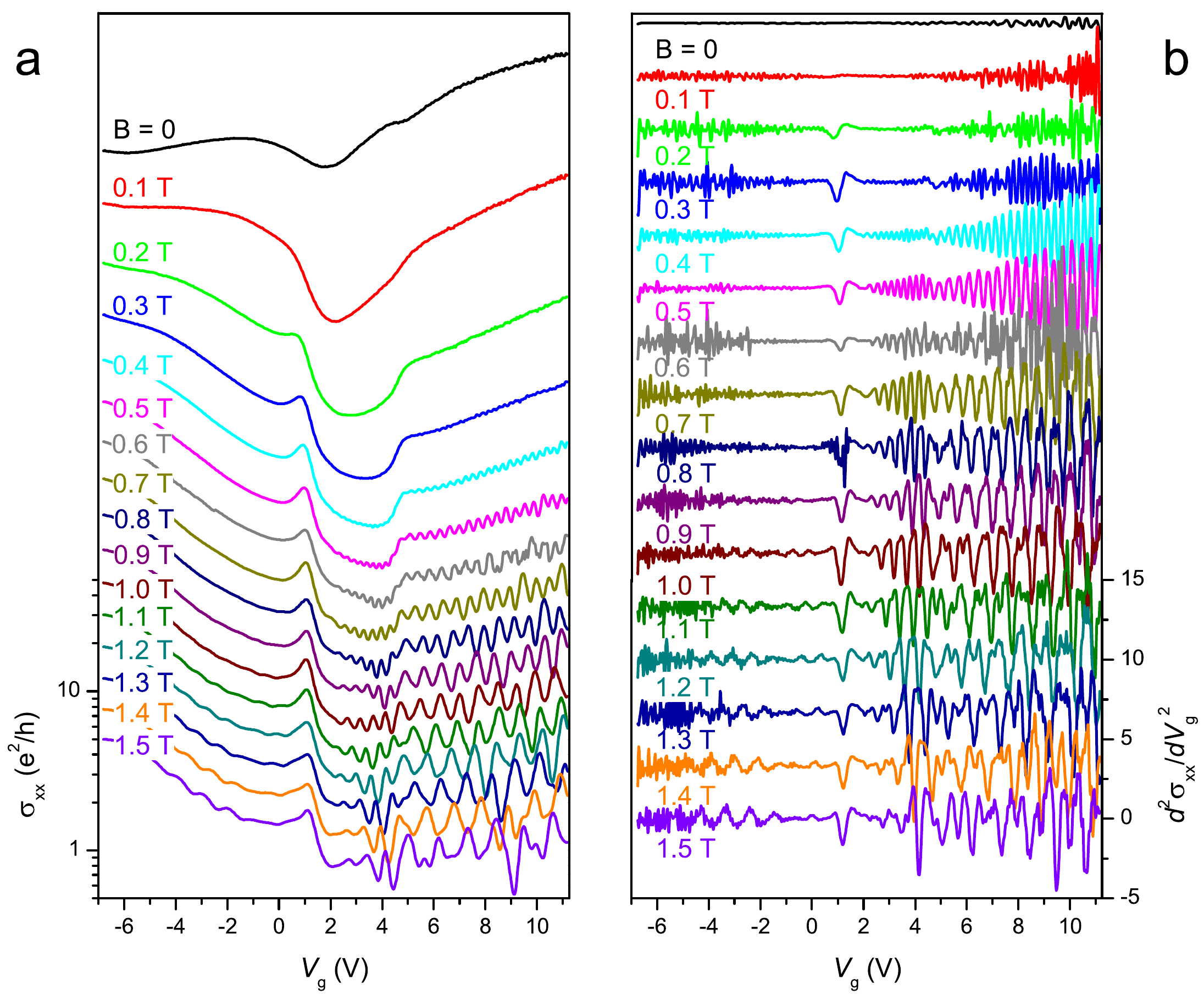}
\caption{Small magnetic field magnetocapacitance data: \textbf{a},
$\sigma_{xx}(V_{\rm g})$ dependencies at magnetic fields from 0
(on the top) to 1.5\,T (bottom). Starting at $B=1.4$\,T each trace was shifted by 0.1\,T on the Y-axis.
 \textbf{b} - The second derivative
$d^2\sigma_{xx}(V_{\rm g})/dV_{\rm g}^2$ of the traces of
panel (a). The SdH oscillations come out more clearly in the second derivative.
Each trace was normalized by its RMS value  and shifted on the Y-axis.}\label{Transport_SmallFields}
\end{figure}

In this section we show some of the raw data and outline how they were prepared for the color maps in the main article.
Fig.~\ref{Cap_Vg}a displays the capacitance as a function of gate voltage $V_{\rm g}$ and for magnetic fields between 0 and 13\,T.
In the color maps we plot the differential magnetocapacitance $\delta C(V_g) =C(V_g)|_{B}-C(V_g)|_{B=0}$, shown for magnetic fields up to 2\,T in Fig.~\ref{Cap_Vg}b. In order to suppress the noise we used standard band-pass Fourier filtering with the cutoff frequencies $f_1 = 0.1-0.3$\,T$^{-1}$ and $f_2 = 1-5$\,T$^{-1}$ depending on the magnetic field range.

The corresponding transport data measured on the same device are plotted in Fig.~\ref{Transport_StrongFields}a. Together with the corresponding $\rho_{xy}(V_g)$ traces (not shown) the resistivity data were converted into $\sigma_{xx}(V_g)$ traces shown in Fig.~\ref{Transport_StrongFields}b. Since the value of $\sigma_{xx}$ varies over several orders of magnitude and in order to improve the clarity of the color maps (Fig.~3d) each $\sigma_{xx}(V_g)$ trace was normalized with respect to its average value using  $\sigma_{xx}^{\rm normalized}(V_g)=  \sigma_{xx}(V_g) / <\sigma_{xx}>$, where $<...>$ means averaging over the whole $V_g$ range. After the normalization procedure the average value of each trace is equal to 1.

The $\sigma_{xx}(V_g)$ data taken at lower magnetic fields are shown in the Fig.~\ref{Transport_SmallFields}a. To work out the oscillations  more clearly we plot the second derivative $d^2\sigma_{xx}(V_{\rm g})/dV_{\rm g}^2$ of the traces, shown in panel a of Fig.~\ref{Transport_SmallFields}b. In order to visualize the low-field oscillations clearly on the color maps each $d^2\sigma_{xx}(V_{\rm g})/dV_{\rm g}^2$ trace was normalized by its root mean square (RMS) value taken over the whole $V_g$ scale. Finally we applied standard low-pass Fourier filtering to remove random noise which is quite pronounced for the lowest field ($B \leq0.3$\,T) traces.



\FloatBarrier

\nocite{}

\bibliography{3dTI}%

\end{document}